\documentclass[12pt,preprint]{aastex}
\newcommand{\etal}{\hbox{ et~al.}}
\newcommand{\eg}{\hbox{e.g.}}

\usepackage{psfig}

\begin{document}

\bibliographystyle{aastex}

%%%%%%%% Title %%%%%%%

\title {Structure of the Accretion Disk in the Lensed Quasar Q~2237+0305 from Multi-Epoch and Multi-Wavelength Narrow Band Photometry}

%%%%%% Author list  %%%%%%%%%%%%%%%%%

\author{J. A. Mu\~noz$^{1,2}$, H. Vives-Arias$^1$, A. M. Mosquera$^{3,4}$, J. Jim\'enez-Vicente$^{5,6}$, C. S. Kochanek$^{3}$, and E. Mediavilla$^{7,8}$}

\bigskip

\affil{$^{1}$Departamento de Astronom\'{\i}a y Astrof\'{\i}sica, Universidad
       de Valencia, E-46100 Burjassot, Valencia, Spain}
\affil{$^{2}$Observatorio Astron\'omico, Universidad de Valencia, E-46980 Paterna, Valencia, Spain}            
\affil{$^{3}$Department of Astronomy, The Ohio State University, %140 West 18th Avenue, 
Columbus, OH 43210, USA}
\affil{$^{4}$ Physics Department, United States Naval Academy, Annapolis, MD 21403, USA}
\affil{$^5$Departamento de F\'{\i}sica Te\'orica y del Cosmos, Universidad de Granada, Campus de Fuentenueva, 18071
Granada, Spain}
\affil{$^6$Instituto Carlos I de F\'{\i}sica Te\'orica y Computacional, Universidad de Granada, 18071 Granada, Spain}      \affil{$^{7}$Instituto de Astrof\'{\i}sica de Canarias, E-38200 La Laguna,
       Santa Cruz de Tenerife, Spain}
\affil{$^{8}$Departamento de Astrof\'{\i}sica, Universidad de La Laguna, E-38200 La Laguna, 
       Santa Cruz de Tenerife, Spain}

%%%%%%%%%%%%%%%% Abstract %%%%%%%%%%%%%%%%%%%%%%%%%%%%

\begin{abstract}

We present estimates for the size and the logarithmic slope of the disk temperature profile of the lensed quasar Q2237+0305, independent of the component velocities. These estimates are based on six epochs of multi-wavelength narrowband images from the Nordic Optical Telescope. For each pair of lensed images and each photometric band, we determine the microlensing amplitude and chromaticity using pre-existing mid-IR photometry to define the baseline for no microlensing magnification. A statistical comparison of the combined microlensing data (6 epochs $\times$ 5 narrow bands $\times$ 6 image pairs) with simulations based on microlensing magnification maps gives Bayesian estimates for the half-light radius of $R_{1/2}=8.5^{+7.5}_{-4.0}\sqrt{ \langle M \rangle/0.3\, M_\odot}$ light-days, and $p=0.95\pm0.33$ for the exponent of the logarithmic temperature profile $T\propto R^{ -1/p}$. This size estimate is in good agreement with most recent studies. Other works based on the study of single microlensing events predict smaller sizes, but could be statistically biased by focusing on high-magnification events.

\end{abstract}


\section{Introduction}\label{sec1}

The basic model for describing the inner regions of quasars is the thin disk model (Shakura \& Sunyaev 1973; Novikov \& Thorne 1973)
which predicts the size of the accretion disk and the radial dependence of its surface temperature. Gravitational microlensing
(Chang \& Refsdal 1979, 1984; see also Kochanek 2004 and Wambsganss 2006) is the main tool used to estimate both parameters, either from time variability or through the wavelength dependence of the microlensing magnification. Microlensing studies
(see e.g. Pooley et al. 2007; Morgan et al. 2010; Sluse et al. 2011, Blackburne et al. 2011, 2014, 2015; Mosquera $\&$ Kochanek 2011;
Jim{\'e}nez-Vicente et al. 2012, 2014; Hainline et al. 2013; Mosquera et al. 2013; MacLeod et al. 2015) have found
that the mean sizes of quasar accretion disks are roughly a factor of 2-3 greater than the predictions of the standard thin disk model.
These differences are too large to be explained by contamination from the broad emission lines and the pseudo-continuum contributions,
or scattering on scales larger than the accretion disk (Dai et al. 2010, Morgan et al. 2010).
Recent measurements of wavelength-dependent continuum lags in two local active galactic nuclei (AGNs) are consistent with the microlensing results
(Shappee et al. 2014, Edelson et al. 2015, Fausnaugh et al. 2015).

The study of the disk temperature profile is more complicated because multi-wavelength observations are needed
to detect chromatic microlensing (Anguita et al. 2008; Poindexter et al. 2008; Bate et al. 2008; Eigenbrod et al. 2008; Floyd et al. 2009;
Blackburne et al. 2011, 2015; Mediavilla et al. 2011a; Mosquera et al. 2011; Mu\~{n}oz et al. 2011; Motta et al. 2012,
Rojas et al. 2014). In this case, it also becomes
more important to separate the contributions from strong emission lines. Such a separation can be done most cleanly using narrow band photometry or spectroscopy.

Many of the studies of the source size in Q~2237+0305 are based on the fitting of its
light curves with tracks extracted from simulated microlensing magnification
maps (Kochanek 2004;  Poindexter \& Kochanek 2010b;  Mosquera et al. 2013). 
As an alternative to light-curve fitting, we will study the size and temperature profile 
of Q~2237+0305 using several epochs of multi-wavelength narrowband observations. This allows us to remove the influence of the broad emission lines on the amplitude of the continuum microlensing (Mosquera et al. 2009). We will follow a
procedure similar to that used in SBS~0909+532 (Mediavilla et al. 2011a) and  HE~1104--1804 (Mu\~noz et al. 2011), but in the case of Q~2237+0305 we have significantly better statistics with six epochs (four will be considered independent) and four quasar images.
This method relies on changes in the microlensing amplitude and chromaticity but not on its dependence with time and hence no velocity estimates are necessary. However, a baseline for the intrinsic flux ratios in the absence of microlensing is needed, and we will define these by the mid-IR flux ratios from Minezaki et al. (2009, see also Agol et al. 2000). In Section 2 we describe the observations and data. Section 3 is devoted to the statistical analysis and in Section 4 we discuss the results.

\section{Observations and Data Analysis}\label{sec2}\

Q~2237+0305 (Huchra et al. 1985) was observed with the 2.56 m Nordic Optical 
Telescope (NOT) located at the Roque de los Muchachos Observatory, La Palma (Spain). We used 
the 2048$\times$2048 ALSFOC detector, with a spatial scale of 0.188 arcsec/pixel. We obtained a total 
of five epochs in 2006 and 2007 which we combine with an earlier epoch from Mosquera et al. (2009) taken in 2003. 
A set of seven narrow filters 
plus the wide Bessel I filter were used, covering the wavelength range 3510-8130 \AA. Table \ref {tab:phot_5d} 
provides a log of our observations. For the first epoch, a V-band image was also taken. In the second 
epoch, the wavelength coverage was poorer, because we were unable to observe with all the filters
due to bad weather conditions. 

The data were reduced using standard IRAF\footnote{IRAF is distributed by the National Optical Astronomy Observatory, 
which is operated by the Association of Universities for Research in Astronomy (AURA) under a cooperative agreement 
with the National Science Foundation} procedures, and pont-spread function (PSF) 
photometry fitting was used to derive the difference in magnitude as a function of wavelength 
between the four images. As in \cite{ana09}, the galaxy bulge was 
modeled with a de Vaucouleurs profile, and the quasar images as point sources. 
This model was convolved with PSFs derived from stars observed in each of the frames, and fit to the 
image using $\chi^2$ statistics following \cite{McLeod1998} and \cite{lehar2000}. Even in the bluest filters, 
the fits were excellent due to the good seeing conditions for most of the epochs (0\farcs6 in I-band). 
Among the narrower filters that we used, only two were affected by the broad emission lines of the quasar. 
The Str\"omgren-u filter 
contains roughly 40\% of the Ly$\alpha$ emission line, based on the SDSS quasar composite spectrum from \cite{vanden01}. And at $\lambda=4110$ \AA, the wavelength range covered by the Str\"omgren-v filter coincides with 
the position and width of the CIV emission line \citep[for more details see][]{ana09}.

\begin{figure}
\begin{center}
\psfig{figure=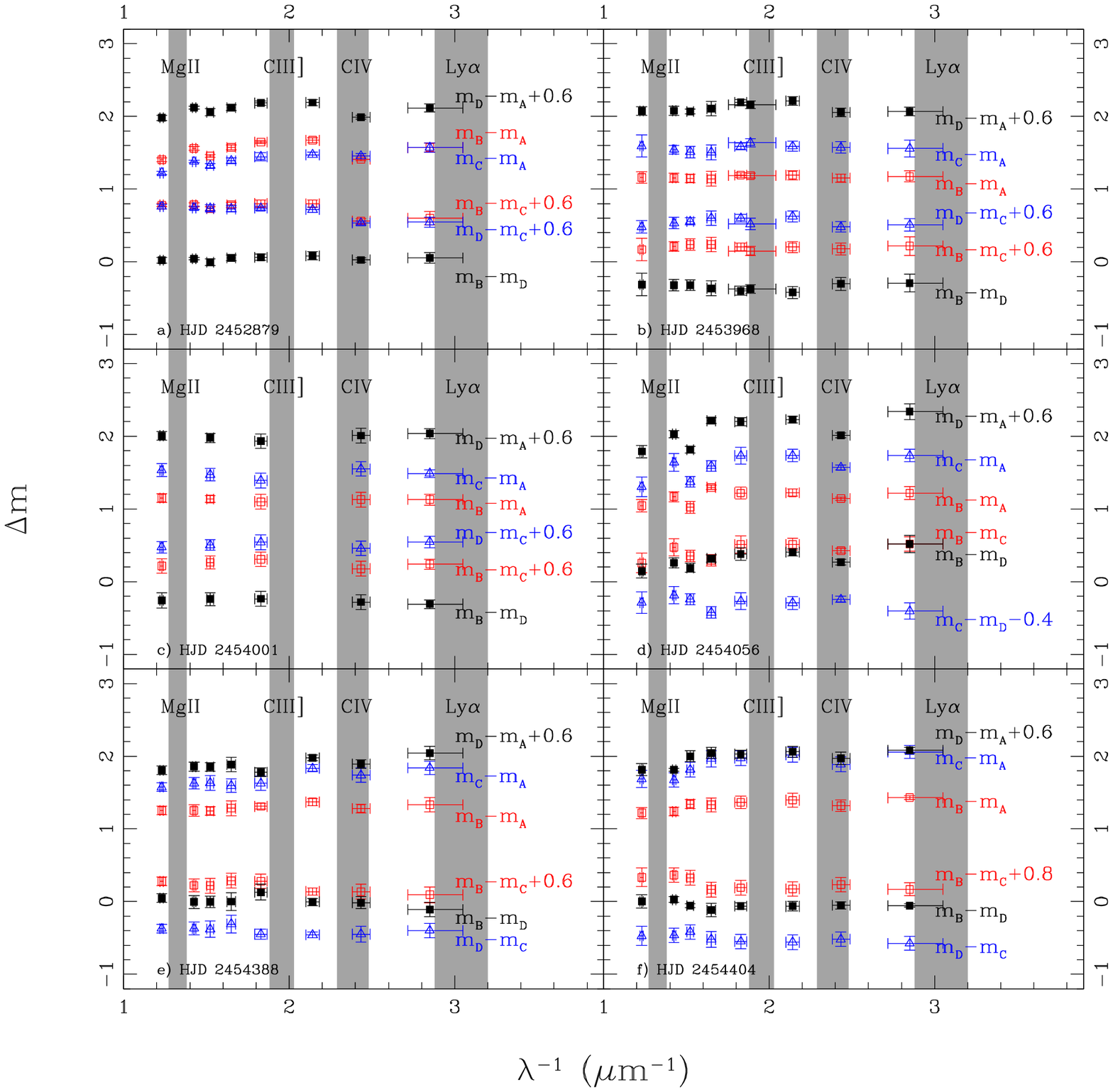,width=16cm}
\caption{\label{mosaico}  Magnitude differences as a function of the
inverse of the observed wavelength for Q~2237+0305 during the six epochs of observation.
Image A showed chromaticity on epoch HJD 2452879 (a), as previously analyzed by Mosquera et al. (2009).
Here we find that image A also shows evidence of chromaticity on HJD 2454056 (d) and
HJD 2454404 (f), as does image B on HJD 2454056 (d). The shaded regions correspond to the wavelength 
location and full width of the most prominent quasar broad emission lines. The horizontal error bars indicate 
the FWHM of the filters.}\end{center}
\end{figure}

Figure \ref{mosaico} shows the results for the six nights and Table 
\ref {tab:phot_5d} reports the photometry for the new data. A clear wavelength dependence of the flux ratios $(m_i-m_A)_{i=B,C,D}$ relative to image $A$ is observed 
for the epochs HJD 2454056 (Fig. \ref{mosaico}d) and 
HJD 2454404 (Fig. \ref{mosaico}f). Chromatic effects are also observed relative to image $B$ $(m_i-m_B)_{i=A,C,D}$ on the first of these two dates.
Since extinction effects would be common to all the epochs (\eg\ Falco\etal\ 1999 or Mu\~noz\etal\ 2004) and the time delays are
negligible, this dependence can only 
be explained by chromatic microlensing of images $A$ and $B$ \citep[for more details see][]{ana09}. Since image $C$ seems little
affected by chromatic microlensing on any of those nights, the difference $(m_C-m_A)^I-(m_C-m_A)^b$ allows us 
to determine the chromaticity in image $A$, with values of $m^{b}_A-m^{I}_A=-0.43\pm0.16$ and 
$m^{b}_A-m^{I}_A=-0.33\pm0.16$ for HJD 2454056 and HJD 2454404 respectively. In a similar way, 
the chromaticity for image $B$ is $m^{b}_B-m^{I}_B=-0.25\pm0.16$ for HJD 2454056. The $I$ and $b$ 
filters were chosen to perform these calculations since they are the reddest and bluest filters that are not affected by the broad 
emission lines. The other epochs do not show any significant dependence on wavelength. In particular the magnitude difference 
as a function of wavelength for the epoch HJD 2454001 (Fig. 1c) is flat given our errors of order $\sim 0.1$ mag. 
This suggests a smaller differential extinction than the estimates of Agol et al. (2000), although we cannot
rule out fortuitous cancellations of color variations due to extinction by chromatic microlensing effects.
Finding chromatic microlensing during this period of observations is not a surprise, since 
the Optical Gravitational Lensing Experiment (OGLE) V-band continuum data \citep{wozniak2000} shows significant brightness variations in the three 
independent magnitude differences at this time. This implies that microlensing is strongly affecting the  quasar images, making chromatic microlensing effects more likely. 

\section{Results}\label{sec3}\

The statistical procedure used to estimate the size of the accretion disk and the strength of the chromaticity 
is a variant of that described in \cite{Mediavillaetal11a}, Mu\~noz et al. (2011) and \cite{Jimenezetal12}.
The main difference is that the observed magnitudes for the four quasar images in the five different bands 
are used simultaneously to calculate the probability distributions.

The magnification maps were calculated using the Inverse Polygon Mapping algorithm
(Mediavilla et al. 2006, Mediavilla et al. 2011a).
We have used canonical values for $\kappa$ and $\gamma$ for
the four images from \cite{Schmidtetal98} and put all the mass into equal mass stars. The models used by \cite{PK1} are very similar but with a mass spectrum for the stars.
The  2000$\times$2000 pixels maps have a pixel size of 0.5 light-days for a mean stellar mass of
$\langle M \rangle = 1M_\sun$ and all linear sizes can be scaled with $(\langle M \rangle/M_\sun)^{1/2}$.
%The half-light radius of the source (related to the Gaussian parameter
 %$r_s$ by $R_{1/2}=1.18r_s$) should be used for comparing to other emission profiles (Mortonson et al. 2005). 
We considered three surface brightness profiles.  In all three models, the scaling of
the source size with wavelength is described by a power-law, $r_s(\lambda_2) = (\lambda_2/\lambda_1)^p r_s(\lambda_1)$
where $\lambda_i$ is the rest wavelength corresponding to each filter.  The first
model we consider is a simple Gaussian, $I(R) \propto \exp(-R^2/2r_s^2)$.  The
second model, $I(R) \propto \left( \exp\left[(R/r_s)^{3/4}\right] -1\right)^{-1}$
becomes the standard thin disk model when $p=4/3$.  The third model,
$I(R) \propto \left( \exp\left[(R/r_s)^{1/p}\right] -1\right)^{-1}$ also becomes
the standard thin disk model when $p=4/3$. The third model corresponds
to a thermally emitting disk with a temperature profile $T \propto R^{-1/p}$
and is well-defined only for $p>1/2$ since we are not including an inner
edge. The second model is a hybrid, but by holding the exponent in the
blackbody function fixed and only varying the scale length, the brightness
profile is well defined for all $p$.  We will refer to the three models as the
Gaussian, hybrid and thin disk models. The length scales $r_s$ will depend
on the profile and $p$, however, 
%we expect the half-light radii of the profiles
%and the estimates of the exponent $p$ (provided it is $>1/2$) to agree based
%on the results of Mortonson et al. (2005).
we expect  the estimates of the half-light radii for the different  profiles
to agree based
on the results of Mortonson et al. (2005).
We use
the bluest observed wavelength, $\lambda_1^{obs}=4670\,\rm \AA$ as the reference wavelength, corresponding to
$\lambda_1=1736$ \AA\ in the rest frame. When no reference is made to a specific
wavelength the size is for this reference wavelength, $r_s\equiv r_s(1736\,\rm \AA)$.

The observed magnitude of image $I=(A,B,C,D)$ is
\begin{equation}
    m_I^{obs}(t_j,i) = m_0(t_j,i) + \mu_I + \delta\mu_I(t_j,i)
\end{equation}
where $m_0(t_j,i)$ is the intrinsic magnitude of the source
at time $t_j$ in filter $i$,
$\mu_I$ is the macro magnification (in mag) of image $I$ and
$\delta\mu_l(t,i)$ is the time varying microlensing magnification (in mag) 
of image $I$ in filter $i$.  Since we have no direct information on the
intrinsic variability of the source or the absolute
magnifications, we ultimately want to work in terms
of magnitude differences between images. We assume
that the mid-IR flux ratios from \cite{Minezakietal09} (cf. their Table 4) correctly estimate
the intrinsic flux ratios $\mu_{IJ}^{ir}$, which means that they
define all the values of $\mu_{IJ}^{ir}=\mu_I - \mu_J$. 
Following Kochanek (2004), we eliminate the intrinsic magnitude of the source $m_0(t_j,i)$ by 
starting from the fit to the four individual images
at a single epoch and wavelength

\begin{equation}
   \chi^2(t_j,i) = \sum_I \sigma_I(t_j,i)^{-2} 
    \left[ m_I^{obs}(t_j,i) - m_0(t_j,i) - \mu_I - \delta\mu_I(t_j,i)\right]^2
\end{equation}

\noindent
and optimizing for the best source model $m_0(t_j,i)$. 
If we then substitute this best fit source model into
$\chi^2(t_j,i)$, we are left with a sum over the six possible pairs between the four images

\begin{equation}
   \chi^2(t_j,i) = \sum_I \sum_{J>I} \sigma_{IJ}(t_j,i)^{-2}
   \left[ \Delta m_I(t_j,i) - \Delta m_J(t_j,i) \right]^2
   \label{eqn:chi2}
\end{equation}
where 
\begin{equation}
   \Delta m_I(t_j,i) = m_I^{obs}(t_j,i) - \mu_I - \delta\mu_I(t_j,i)
\end{equation}

\noindent
and the errors $\sigma_{IJ}(t_j,i)$ are defined in the 
equation 7 of Kochanek (2004) but
reduce to $\sigma_{IJ}(t_j,i) = 2\, \sigma(t_j,i)$ 
 if $\sigma_I=\sigma_J(\equiv \sigma)$ as we will
assume here\footnote{Although there are slightly different errors for every individual image we have chosen to use  the average measurement error $\sigma=0.08$
 for weighting all the data.}. This expression only depends on
$\mu_I-\mu_J$, whose values are supplied by the
mid-IR flux ratios.  Thus, the full $\chi_j^2 $ for a given epoch $t_j$
becomes the sum of Equation~\ref{eqn:chi2}
over the filters,
\begin{equation}
   \chi_j^2 =  \sum_i \chi^2 (t_j,i)
\end{equation}
which can be evaluated for any microlensing trial 
defining the values of $\delta\mu_I(t_j,i)$.

For a given pair of parameters 
$(r_s,p)$, the magnification maps for the four images $A,B,C,D$ are convolved to the size $r_s(\lambda/\lambda_0)^p$ appropriate for each wavelength and 
the $\chi_j^2$ are calculated for N=$10^8$  randomly selected locations in each of the four magnification maps. The probability
density function $P_j(r_s,p)$ for each epoch is computed as the sum

\begin{equation}
P_j(r_s,p)\propto
\sum_{i=1}^N{{{e^{-{1\over 2}\chi_j^2}}}}
\end{equation}

\noindent
over a 2D grid of values for $r_s$ and $p$. We have used a logarithmic grid in $r_s$ such that
$\ln (r_s^i/\mbox{light-days})=0.3\times i$ for $i=0, \cdots,17$ and a linear grid in $p$ with $p^j=0.25\times j$ for $j=0,\cdots,9$.
This way, $r_s$ spans from roughly 1 to 165 light-days and $p$ runs from 0 to 2.25. For the thin disk model,
which is only defined for  $p>1/2$, we ran extra cases at $p=0.55$ and $0.68$ and then follow the remainder of the sequence.

We have data for six different epochs between August 2003 and October 2007.
Some of the epochs are very close on time and we have combined them into a single epoch because they cannot be considered independent. Epoch HJD 2453968 (Aug 21st 2006) is combined with HJD 2454001 (Sep 23rd 2006) and epoch HJD 2454388 (Oct 15th 2007) is
combined with HJD 2454404 (Oct 31st 2007). The dispersion around the mean for all values of the combined epochs is 0.08 mags, which is the same as the measurement error we adopted for the merit function above.

The joint probability distribution ${\cal P}(r_s,p)$ for the four independent epochs is obtained by the product of the probabilities for each of the epochs

\begin{equation}
{\cal P}(r_s,p)=\prod_j P_j(r_s,p)
\end{equation}

\noindent
The resulting joint probability distributions of the
three surface brightness profiles are shown in Figure 2 and are compared
to previous size estimates for Q~2237+0305 in Figure 3. The maximum likelihood estimate corresponds to $r_s=(15.6^{+5.5}_{-7.9}, 5.3^{+1.9}_{-2.4}, 27^{+22}_{-19} )$ $\sqrt{\langle M \rangle/0.3M_\odot}$ light-days and $p=(0.50^{+0.15}_{-0.27}, 1.0^{+0.25}_{-0.50}, 0.85^{+0.25}_{-0.25})$  for the Gaussian, hybrid and thin disk models respectively. Using a logarithmic (linear) prior for $r_s$ ($p$), we obtain  Bayesian estimates for the expected values of $r_s=(7.0^{+10.0}_{-4.1}, 3.7_{-1.8}^{+3.5}, 7.9_{-5.3}^{+16.6}) \sqrt{\langle M \rangle/0.3M_\odot}$ light-days and $p=(0.66\pm 0.32, 1.04\pm0.29, 0.95\pm0.33)$.
As a test we also computed the joint probability distribution without merging the closely separated
epochs and obtained very similar results.

\section{Discussion and Conclusions}\label{sec4}

At $\lambda_1=1736\ $\AA, the half-light radius estimates for the three (Gaussian, hybrid, and thin disk) models of $R_{1/2}=(8.3^{+11.8}_{-4.8}, 9.0^{+8.4}_{-4.4}, 8.5_{-4.0}^{+7.5}) \sqrt{\langle M \rangle/0.3M_\odot}$ light-days  are remarkably similar (see Figure 2), consistent with
the prediction from Mortonson et al. (2005) that estimates of the half-light radius should be
 insensitive to the particular surface brightness profile shape.
Our $R_{1/2}$ estimations are also in agreement with previous estimates by Poindexter \& Kochanek (2010b; $R_{1/2}=5.4\pm3.2$ light-days), Sluse et al. (2011; $R_{1/2}=3.4^{+6.4}_{-2.4}$ light-days) and Mosquera et al.  (2013; $R_{1/2}=9.9^{+5.1}_{-3.3}$ light-days), as shown in Figure \ref{pdfh}. The latter two results are also scaled
to $\langle M \rangle=0.3M_\odot$. The Poindexter \& Kochanek (2010b) result marginalizes over the uncertainties in
$\langle M \rangle$, finding $\langle M \rangle=0.52 M_\odot$ ($0.12  M_\odot <M < 1.94 M_\odot$).
Poindexter \& Kochanek (2010ab), Mosquera et al. (2013) and Sluse et al. (2011) set the scales by adopting priors on the
effective source velocity rather than choosing a fixed mean mass $\langle M \rangle$. That our results agree demonstrates 
that these priors are reasonable.

Analyses of individual high magnification events tend to measure smaller sizes. For example Anguita
et al. (2008) find $R_{1/2}=1.0^{+0.2}_{-0.5}$ light-days and Eigenbrod et al. (2008) find $R_{1/2}=3.0\pm 2.0$ light-days. However, studies of high magnification events are likely biased toward small quasar size estimates because it is easier
to obtain high magnifications with small sources (e.g. Kochanek 2004; Eigenbrod et al. 2008; Blackburne et al. 2011). It is also interesting to note that without considering any velocity prior  (i.e. adopting a uniform prior), the size determinations of  Eigenbrod et al. (2008) and Anguita et al. (2008) increase by a factor $\sim 4$ (see Sluse et al. 2011) but would then require cosmologically
unrealistic peculiar velocities for the lens/source.

In any case, most of the results derived from the optical may be reconciled near a value of  $\sim (7\pm 4) \sqrt{\langle M \rangle/ 0.3 M_\odot}$ light-days. This value is large compared to the predictions of the thin disk model based on the flux ($\sim 1$ 
light-day, Mosquera $\&$ Kochanek 2011). It is in agreement with the results from other lenses, where the Morgan et al. (2010) black hole-mass-size correlation predicts a size of $\sim 5$ light-days for an
estimated black hole mass of $1.2\times 10^9 M_\odot$ (Assef et al. 2011). This size discrepancy is now also seen in the
recent size estimates for two local AGN using measurements of continuum lags 
(Shappee et al. 2014, Edelson et al. 2015, Fausnaugh et al. 2015). The size problem is clearly not unique to the 
microlensing method.

The hybrid and thin disk models both find $p=1.0\pm0.3$ for the slope of the dependence of the disk size on wavelength, while the Gaussian model favors a steeper (in temperature) slope of $p=0.7\pm0.3$. The three 
estimates are mutually compatible and smaller than the prediction of the standard thin disk model ($p=4/3$). 
The differences for the (presumably) more physical hybrid or thin disk models are small enough to represent
only a modest inconsistency. Experiments with broader wavelength ranges are needed to provide better estimates of the temperatures exponent. 

%This result has been found in several lensed quasars by other authors (see Jim{\'e}nez-Vicente et al. 2014 and references therein). Such a steeper temperature profile ($T\propto R^{ -1/p}$ ) exacerbates the disk size problem (see Morgan et al. 2010).

\bigskip

\noindent Acknowledgments:
This research was supported by the Spanish Ministerio de Educaci\'on y Ciencia with the grants
 AYA2010-21741-C03-01/02,  AYA2011-24728, AYA2013-47744-C3-1/3-P and AYA2014-53506-P.
 JAM is also supported by the Generalitat
Valenciana with the grant PROMETEO/2014/60. JJV is also supported by the
Junta de Andaluc\'{\i}a through the FQM-108 project. 
AMM  is supported by NSF grant AST-1211146.

%%% REFERENCES %%%%%

%%% FIGURES %%%
\clearpage

\clearpage

\begin{figure}
\begin{center}
\psfig{figure=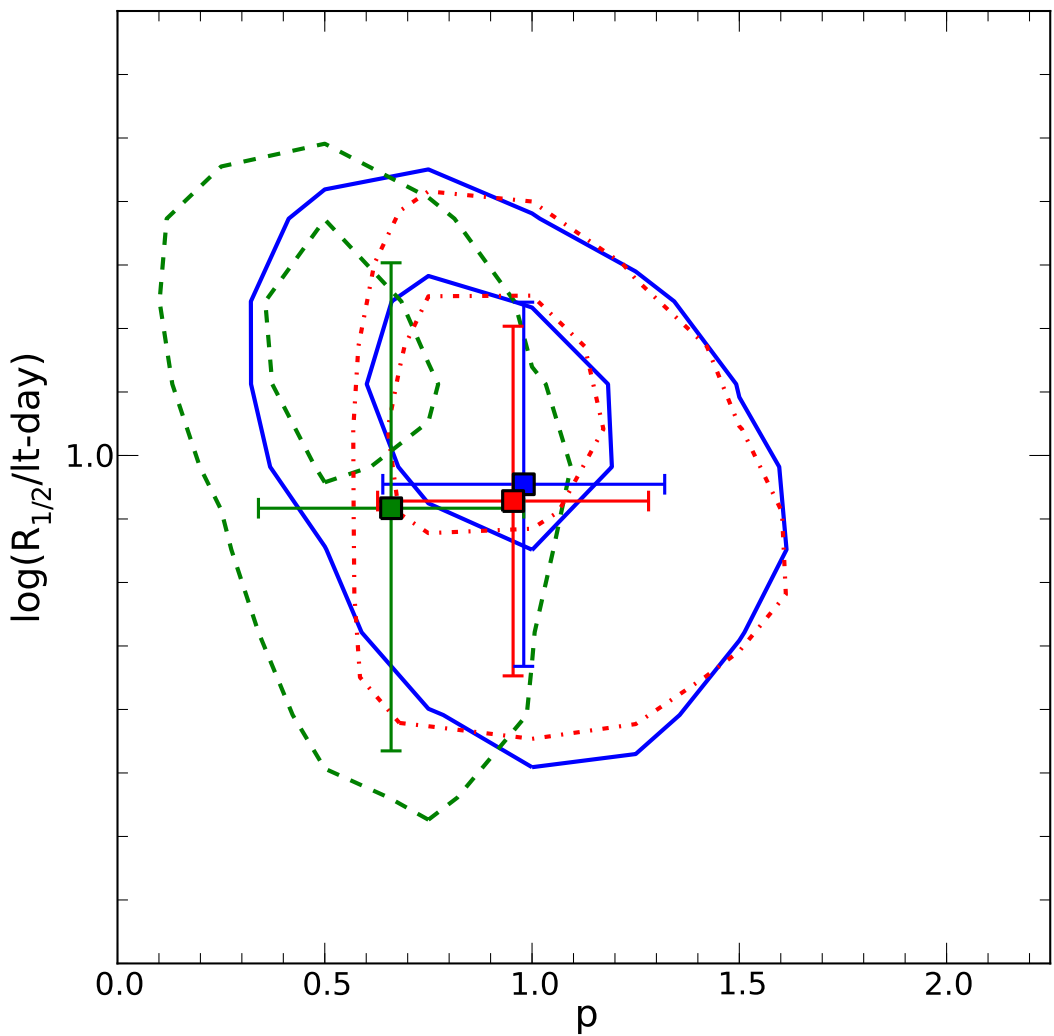,width=16cm}
\end{center}
\caption{\label{pdf3} Joint probability density function ${\cal P}(r_s,p)$ in terms of the half-light radius 
$R_{1/2}$, at rest $\lambda_1=1736$ \AA, and the  logarithmic  slope $p$ ($r_s\propto \lambda^p$).
The  contours  for each model correspond to the 1-$\sigma$ and 
2-$\sigma$ levels for one parameter. 
The Gaussian, hybrid and thin disk models are shown by the green (dashed), blue (solid) and dotted (red) colors (lines).
The filled squares are the Bayesian estimates for the
expected values of $R_{1/2}$ and $p$. }

\end{figure}

\begin{figure}
\begin{center}
\psfig{figure=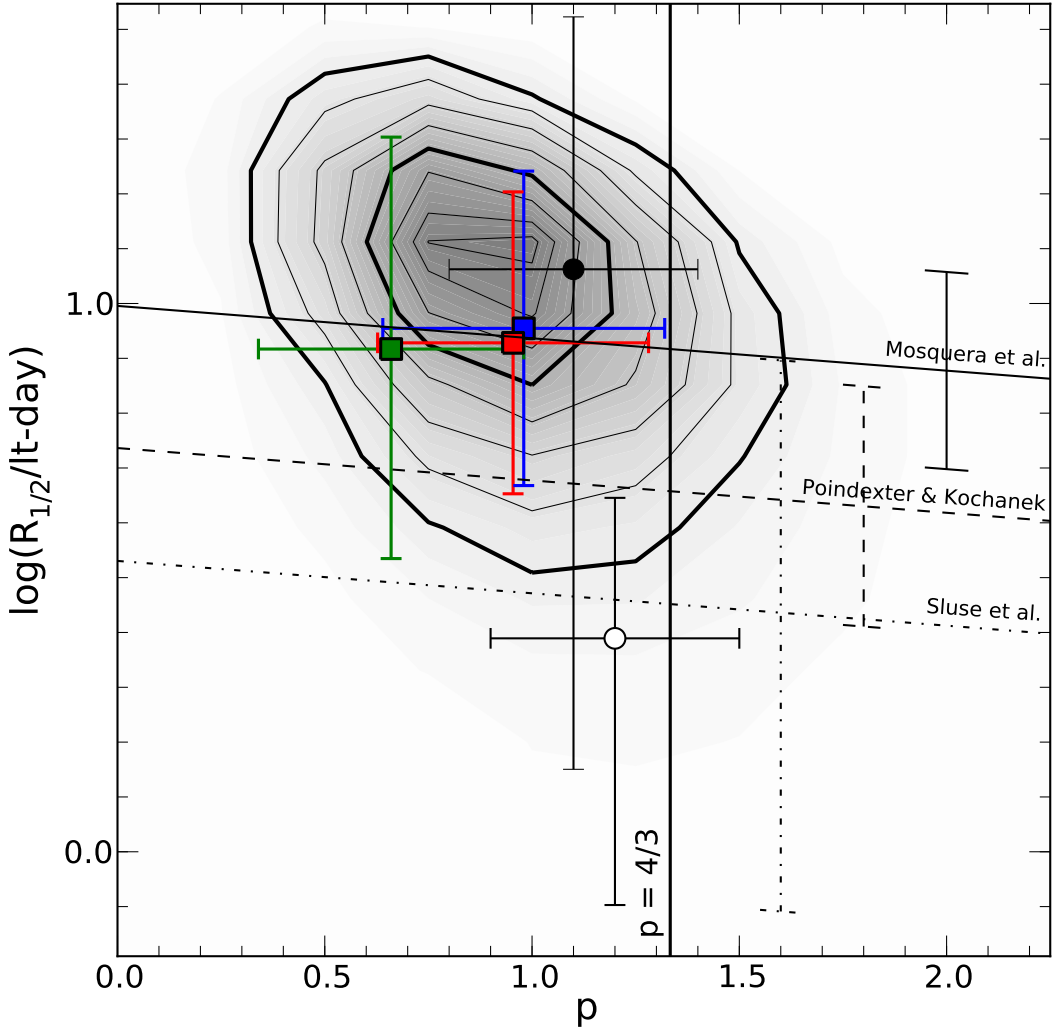,width=16cm}
\end{center}
\vspace*{-1cm}
\caption{\label{pdfh} Joint probability density function ${\cal P}(r_s,p)$ in terms of the half-light radius 
$R_{1/2}$, at rest $\lambda_1=1736$ \AA,  and the logarithmic  slope $p$ ($r_s\propto \lambda^p$) for the hybrid model.
% profile $I(R)\propto 1/(\exp[(R/r_s)^{3/4}]-1)$.
The separation between consecutive contours corresponds to 0.25 $\sigma$, the 1-$\sigma$ and 
2-$\sigma$ for one parameter contours are heavier. The blue filled square is our Bayesian estimate for the
expected value of $R_{1/2}$ and $p$ for the hybrid model (the values for the Gaussian and thin disk models are also plotted as green and red filled squares respectively for comparison). The open (filled) circle corresponds to the measurement by Eigenbrod et al (2008) with (without) a velocity prior. Straight lines correspond to the measurements by  Poindexter \& Kochanek 2010b (dashed line), Sluse et al. 2011 (dotted-dashed line) and Mosquera et al. 2013 (continuous line), that have no estimate on $p$; the associated error bars correspond to their $\pm 1\sigma$  uncertainties.
The size comparisons have been made setting the mean mass of the stars to $\langle M \rangle= 0.3\,M_\odot$.}
\end{figure}

%%% TABLES%%%

%\begin {document} 
\begin{deluxetable}{rrrrr} 
\tabletypesize{\small}
\tablecaption{\label{tab:phot_5d} Q~2237+0305 PHOTOMETRY}
\tablewidth{0pt}
\tablehead{
\colhead{Filter} & \colhead{$m_B-m_A$} & \colhead{$m_C-m_A$} & \colhead{$m_D-m_A$}&\colhead{Observation date}}
\startdata
Str-u ($\lambda=3510$ \AA) & 1.18$\pm$0.07 & 1.56$\pm$0.11 & 1.47$\pm$0.06 & Aug 21 2006 \\
Str-v ($\lambda=4110$ \AA) &1.15$\pm$0.06 & 1.58$\pm$0.08 & 1.46$\pm$0.06 & Aug 21 2006 \\
Str-b ($\lambda=4670$ \AA) & 1.19$\pm$0.06 & 1.59$\pm$0.07 & 1.61$\pm$0.06 & Aug 21 2006 \\
V-band ($\lambda=5300$ \AA) &  1.19$\pm$0.04 & 1.64$\pm$0.06 & 1.56$\pm$0.06 & Aug 21 2006 \\
Str-y ($\lambda=5470$ \AA) & 1.19$\pm$0.03 & 1.59$\pm$0.05 & 1.59$\pm$0.04 & Aug 21 2006 \\
Iac\#28 ($\lambda=6062$ \AA) & 1.14$\pm$0.10& 1.50$\pm$0.10 & 1.51$\pm$0.10 & Aug 21 2006 \\
H$\alpha$ ($\lambda=6567$ \AA) & 1.15$\pm$0.06 & 1.51$\pm$0.08 & 1.47$\pm$0.03 & Aug 21 2006 \\
Iac\#29 ($\lambda=7015$ \AA) &  1.16$\pm$0.06 & 1.54$\pm$0.06 & 1.48$\pm$0.06 & Aug 21 2006\\
I-band ($\lambda=8130$ \AA) &1.16$\pm$0.08 & 1.59$\pm$0.15 &1.48$\pm$0.06  & Aug 21 2006\\
Str-u ($\lambda=3510$ \AA)& 1.13$\pm$0.08 &1.49$\pm$0.05 & 1.44$\pm$0.07 & Sep 23 2006 \\
Str-v ($\lambda=4110$ \AA) &1.13$\pm$0.10 &1.55$\pm$0.10 & 1.41$\pm$0.10 &Sep 23 2006 \\
Str-y ($\lambda=5470$ \AA) & 1.10$\pm$0.10&1.39$\pm$0.10 &1.34$\pm$0.10 & Sep 23 2006\\
H$\alpha$ ($\lambda=6567$ \AA) &1.14$\pm$0.05 &1.47$\pm$0.08 & 1.38$\pm$0.06 &Sep 23 2006 \\
I-band ($\lambda=8130$ \AA) & 1.15$\pm$0.06 & 1.53$\pm$0.09 & 1.41$\pm$0.06  &Sep 23 2006\\
Str-u ($\lambda=3510$ \AA)&  1.22$\pm$0.09& 1.74$\pm$0.08 &  1.74$\pm$0.10 & Nov 17 2006\\
Str-v ($\lambda=4110$ \AA) & 1.14$\pm$0.02 & 1.57$\pm$0.03 & 1.41$\pm$0.03 & Nov 17 2006\\
Str-b ($\lambda=4670$ \AA) & 1.22$\pm$0.05 &  1.74$\pm$0.09 &  1.63$\pm$0.05 & Nov 17 2006\\
Str-y ($\lambda=5470$ \AA) & 1.22$\pm$0.08 & 1.73$\pm$0.11 & 1.60$\pm$0.06 & Nov 17 2006\\
Iac\#28 ($\lambda=6062$ \AA) & 1.30$\pm$0.03 &  1.59$\pm$0.07 & 1.62$\pm$0.04 &Nov 17 2006 \\
H$\alpha$ ($\lambda=6567$ \AA) & 1.02$\pm$0.08 & 1.38$\pm$0.07 & 1.22$\pm$0.04 & Nov 17 2006\\
Iac\#29 ($\lambda=7015$ \AA) & 1.17$\pm$0.07 & 1.64$\pm$0.12 & 1.43$\pm$0.05 &Nov 17 2006\\
I-band ($\lambda=8130$ \AA) & 1.05$\pm$0.08 &  1.31$\pm$0.13 & 1.19$\pm$0.08  &Nov 17 2006\\
Str-u ($\lambda=3510$ \AA)& 1.33$\pm$0.10  &  1.84$\pm$0.09 &1.44$\pm$0.09  &  Oct 15 2007\\
Str-v ($\lambda=4110$ \AA) & 1.28$\pm$0.06 & 1.74$\pm$0.10 & 1.29$\pm$0.06 & Oct 15 2007 \\
Str-b ($\lambda=4670$ \AA) & 1.37$\pm$0.05 & 1.83$\pm$0.05  & 1.37$\pm$0.04 & Oct 15 2007\\
Str-y ($\lambda=5470$ \AA) & 1.31$\pm$0.04 & 1.63$\pm$0.10 &  1.18$\pm$0.06 &  Oct 15 2007 \\
Iac\#28 ($\lambda=6062$ \AA) & 1.28$\pm$0.10 & 1.60$\pm$0.09 &  1.29$\pm$0.10  & Oct 15 2007\\
H$\alpha$ ($\lambda=6567$ \AA) & 1.24$\pm$0.06 & 1.63$\pm$0.10 & 1.25$\pm$0.06  & Oct 15 2007 \\
Iac\#29 ($\lambda=7015$ \AA) & 1.25$\pm$0.08 & 1.63$\pm$0.08 & 1.26$\pm$0.06 & Oct 15 2007\\
I-band ($\lambda=8130$ \AA) & 1.25$\pm$0.06& 1.58$\pm$0.06 & 1.21$\pm$0.06 & Oct 15 2007\\
Str-u ($\lambda=3510$ \AA)& 1.43$\pm$0.03 &2.06$\pm$0.09 & 1.48$\pm$0.04 & Oct 31 2007\\
Str-v ($\lambda=4110$ \AA) & 1.32$\pm$0.08 &1.89$\pm$0.10 & 1.37$\pm$0.09 &Oct 31 2007\\
Str-b ($\lambda=4670$ \AA) & 1.39$\pm$0.09 & 2.02$\pm$0.10 & 1.46$\pm$0.07 & Oct 31 2007\\
Str-y ($\lambda=5470$ \AA) & 1.37$\pm$0.08 & 1.98$\pm$0.10 & 1.43$\pm$0.06 & Oct 31 2007\\
Iac\#28 ($\lambda=6062$ \AA) & 1.33$\pm$0.09 &1.97$\pm$0.11 & 1.45$\pm$0.08& Oct 31 2007\\
H$\alpha$ ($\lambda=6567$ \AA) &1.34$\pm$0.06 &  1.82$\pm$0.10  &  1.40$\pm$0.08 & Oct 31 2007\\
Iac\#29 ($\lambda=7015$ \AA) &  1.24$\pm$0.06& 1.67$\pm$0.10 &  1.21$\pm$0.02 & Oct 31 2007\\
I-band ($\lambda=8130$ \AA) &  1.22$\pm$0.08 & 1.69$\pm$0.12 & 1.22$\pm$0.08 & Oct 31 2007\\
\enddata
\end{deluxetable}
%\end{document}

\end{document}